\documentclass[conference]{IEEEtran}
\IEEEoverridecommandlockouts
\usepackage{amsmath,amsfonts}
\usepackage{algorithmic}
\usepackage{algorithm}
\usepackage{array}
\usepackage{textcomp}
\usepackage{stfloats}
\usepackage{url}
\usepackage{verbatim}
\usepackage{graphicx}
\usepackage{caption}
\usepackage{romannum}
\usepackage{cite}
\usepackage{amsmath}
\usepackage{makecell}
\usepackage{multirow}
\usepackage{hyperref}
\usepackage{xcolor}
\usepackage{subfigure}

\begin{document}

\title{Thermal-Guided Ensemble Reinforcement Learning for Regime-Adaptive Short-Term GPU Power Forecasting in AI Data Centers

\thanks{This research was supported by the National Science Foundation through award number 2346213 and 2418359.}
}

\author{
\IEEEauthorblockN{Yuanshuo Zhang, Ying Zhang, Ziying Wang}
\IEEEauthorblockA{
\textit{School of Electrical and Computer Engineering}\\
Oklahoma State University\\
Stillwater, OK, USA\\
\{yuanshuo.zhang, y.zhang, zy.wang\}@okstate.edu}
\and
\IEEEauthorblockN{Fanping Sui}
\IEEEauthorblockA{
\textit{Department of Mechanical Engineering}\\
University of California Berkeley\\
Berkeley, CA, USA\\
fpsui@berkeley.edu
}
}

\maketitle                                               
\begin{abstract}
AI data centers (AIDCs) introduce bursty and regime-dependent load patterns, posing significant challenges for reliable energy management and predictive grid control. To address these challenges, this letter proposes a thermal-guided ensemble reinforcement learning framework for probabilistic AIDC load forecasting. A soft actor-critic (SAC) agent with a simplex-preserving policy adaptively weights LSTM, TCN, and Transformer predictors at every time step based on their realized errors. A thermal-guided neural correction estimates the ensemble residual, while a sensitivity-projection regularizer enforces the non-negative sensitivity of power to GPU utilization and temperature. The corrected ensemble output is further processed by a Bayesian probabilistic layer to produce predictive distributions and quantify forecasting uncertainty. 
Case studies on 1-second MIT Supercloud telemetry show that the proposed method reduces the normalized mean absolute error to about 1\% across all computing regimes, compared with 4--11\% for deep learning-based and ensemble learning benchmarks, and converges faster with narrower prediction intervals than a DRL ensemble without thermal guidance.
\end{abstract}

\begin{IEEEkeywords}
AI data center, load forecasting, probabilistic forecasting, reinforcement learning, physics-informed neural networks, uncertainty, thermal.
\end{IEEEkeywords}

\section{Introduction}
\IEEEPARstart{T}{he} rapid growth of AI data centers (AIDCs) has led to unprecedented workload dynamics driven by large-scale GPU clusters and training job scheduling \cite{sheng2026power,chen2025electricity}. AIDC power can swing by a large fraction of the facility rating within seconds when synchronized training jobs start, checkpoint, or terminate \cite{choukse2025power}. Such sharp power ramps, if not predicted and controlled, increase the risk of oscillations and grid instability \cite{abudyak2026mitigating,kwon2025operational}. Moreover, AIDCs are increasingly regarded as grid-interactive flexible assets \cite{colangelo2025ai}, which requires anticipating the power trajectory. Therefore, reliable short-term load forecasting at the second level is essential for flexible job scheduling, thermal management, and secure grid operation.

Despite extensive research in deep learning-base load forecasting, \cite{mu2025pesgm} reveals that
rapid transitions among these regimes significantly degrade the robustness of existing forecasting models,
such as long short-term memory (LSTM). Large errors, once aggregated across hundreds of computing nodes in data centers for grid integration, pose high risks to grid reliability and operational stability. However, \cite{mu2025pesgm} reveals that rapid transitions among idle, busy, and ramping regimes significantly degrade the robustness of such models when they are applied to AIDC loads, and that no single architecture dominates in all regimes. To this end, an adaptive ensemble learning is developed in \cite{Wang2026RegimeAdaptive} to combine complementary predictors, yet fixed or validation-tuned weights cannot follow abrupt regime switches. Deep reinforcement learning (DRL) is a natural tool for such sequential decisions, and off-policy maximum-entropy agents have shown robust and sample-efficient adaptation in power system control \cite{zhang2023off,9916943}.
Recent studies have shown that AIDC operation exhibits strong physical
coupling among workload dynamics, power draw, and thermal behavior
\cite{ji2026sensitivity}. \cite{saleh2026physics}
introduces a physics-aware 1-minute GPU power
forecasting model, where the residuals from the thermal dynamics of a
coupled two-node thermal resistance-capacitance network are used to penalize discrepancies
between the observed temperature dynamics of the GPU die and memory
and those implied by the predicted power. 
However, existing forecasting methods neglect such intrinsic coupling between load and thermal behavior. To this end, leveraging this coupling offers an opportunity to develop a regime-adaptive predictor that effectively anticipates computing transitions.

To fill this gap, this paper proposes a thermally guided ensemble reinforcement learning (TERL) approach for probabilistic AIDC load forecasting. 
A soft actor-critic (SAC)-based ensemble learning framework is proposed to automatically weight selected base predictors that adapts to various computing regimes while preserving the positive power–utilization and
power–temperature correlation. Furthermore, the proposed framework incorporates a thermally guided neural correction module that estimates the forecasting residual and adds it back to the baseline prediction. The corrected representation is then passed through a Bayesian output layer to generate probabilistic forecasts and quantify predictive uncertainty. The main contribution of this paper is summarized below: 

\begin{enumerate}
\item A thermal-guided adaptive ensemble DRL framework  is proposed for 1-second GPU power forecasting, which  utilizes the non-negative power sensitivities to GPU utilization and temperature \cite{ji2026sensitivity}. 
The resulting thermal-consistency term enters the DRL reward, and the ensemble policy is thus steered toward physically consistent forecasts.
\item A Bayesian output layer with a variational weight posterior is integrated, and its negative evidence-lower-bound loss is used as the reward. This unifies probabilistic calibration, physical consistency, and regime adaptation in a loss function. Comparative case studies with individual predictors and ensemble methods show that the framework significantly improves second-level forecasting accuracy while generating thermal-calibrated probabilistic forecasts with explicit uncertainty bounds. 
\end{enumerate}

\section{Ensemble Learning Framework for AIDC Power Draw Forecasting}

Given an an AIDC cluster, the feature vector at time step $t$ is defined as \cite{mu2025pesgm}:
\begin{equation}
    \mathbf{x}_t =
\left[
u_t^{\mathrm{gpu}},
u_t^{\mathrm{mem}},
T_t^{\mathrm{gpu}},
T_t^{\mathrm{mem}},
n_t^{\mathrm{job}}
\right]
\end{equation}
where $u_t^{\mathrm{gpu}}$ and $u_t^{\mathrm{mem}}$ denote GPU and memory utilization, $T_t^{\mathrm{gpu}}$ and $T_t^{\mathrm{mem}}$ denote GPU and memory temperatures, and $n_t^{\mathrm{job}}$ is the number of active jobs.

Short-term forecasting aims at mapping the past $L$-length input sequence $\mathbf{X}_t = \mathbf{x}_{t-L+1:t}$ to the GPU power $y_{t+h}$, where $h$ is the forecasting horizon. Ensemble learning techniques are used when a single method cannot always perform effectively. 
To capture diverse load dynamics, LSTM, temporal convolutional network (TCN), and Transformer are adopted as $N=3$ base models, owing to their specialization in sequential, burst, and long-range temporal patterns, respectively \cite{mu2025pesgm}; the case study validates this observation. The GPU power at $t$ is predicted by the $n$th base predictor:
\begin{equation}
\hat{y}_{t}^{(n)}
=
f_n(\mathbf{X}_t;\Theta_n),
\quad \forall n
\end{equation}
where 
$\Theta_n$ denotes the trainable parameters of base learner $n$. 

The ensemble prediction is formulated as a weighted combination of $\hat{y}_{t}^{(n)}$ through non-negative weight $w_t^{(n)}$ assigned to the $n$-th base learner as:
\begin{equation}\label{3}
\hat{y}_{t}^{\mathrm{ens}}
=
\sum_{n=1}^{N}
w_t^{(n)}
\hat{y}_{t}^{(n)} \quad \text{with} \sum_{n=1}^{N} w_t^{(n)} = 1
\end{equation}

Fixed ensemble weights may fail during abrupt AI workload transitions, and their non-adaptive optimization hinders rapid adaptation across all scenarios. In the next section, a novel DRL-based method is proposed for adaptive ensemble of each base learner by automatically assigning $w_t^{(n)}$.

\begin{figure}[!t]
    \centering
        \includegraphics[width=0.49\textwidth]{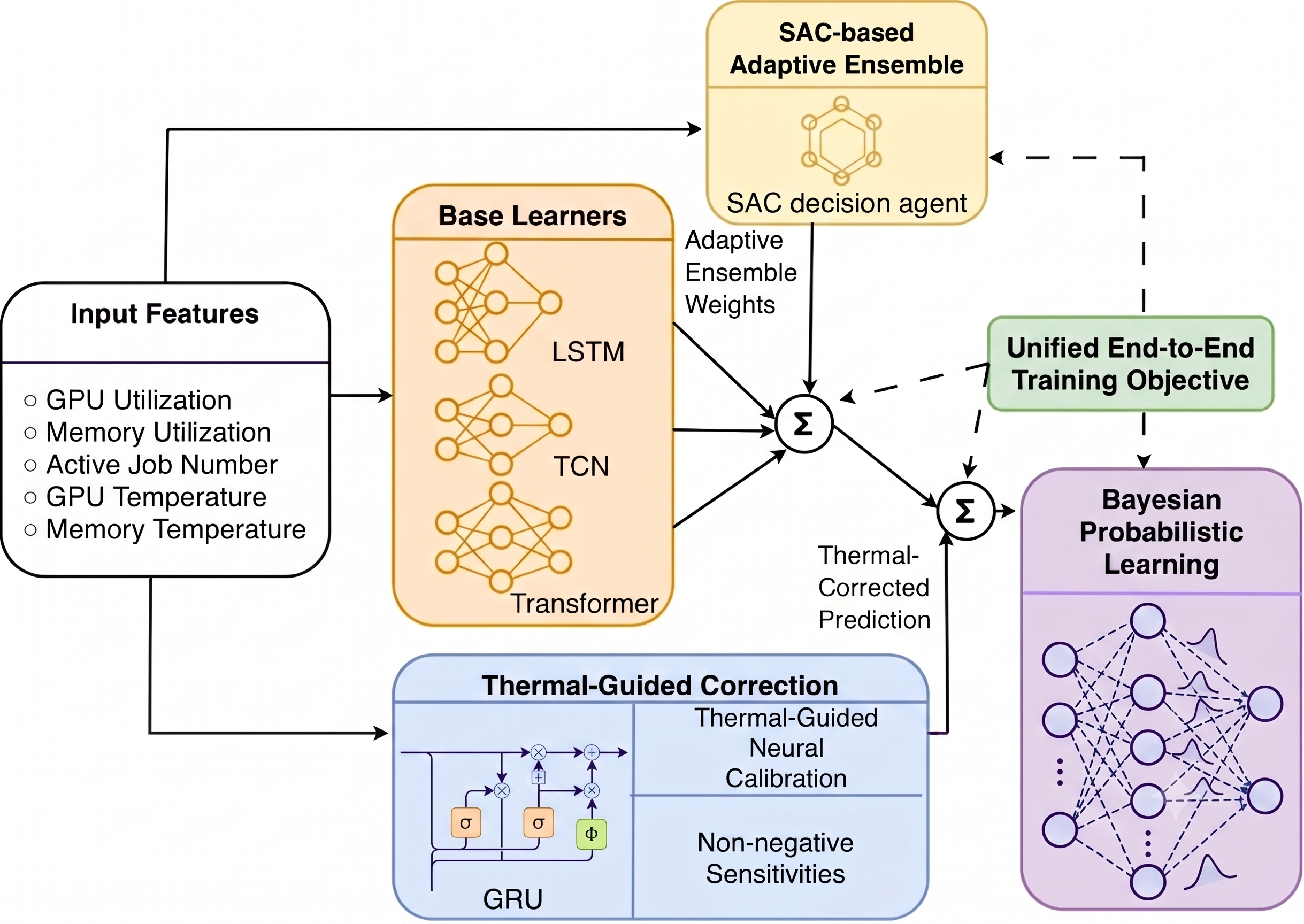}
    \vspace{-5pt}
    \caption{Overview of proposed thermal-guided ensemble reinforcement learning for AIDC load forecasting}
    \label{fig:framework}
\end{figure}
\section{Proposed Regime-Adaptive Thermal-Guided RL for Probabilistic AIDC Load Forecasting}
The proposed TERL algorithm is shown in Fig.~\ref{fig:framework}. Since purely data-driven forecasting may violate thermal--power relationships, a thermal-guided correction and a Bayesian output layer are integrated into the DRL-based ensemble and tied together through the reward.

\subsection{SAC-Based Regime-Adaptive Ensemble Weighting}
The weight selection is modeled as a Markov decision process. The state is $\mathbf{s}_t=[\mathrm{vec}(\mathbf{X}_t),\mathbf{e}_t]$, where $\mathbf{e}_t=[e_t^{(1)},\dots,e_t^{(N)}]$ and
\begin{equation}\label{eq:err}
e_t^{(n)} = y_t - f_n(\mathbf{X}_{t-h};\Theta_n)
\end{equation}
is the most recently \emph{realized} error of the $n$th learner, i.e., the error of the forecast issued $h$ steps earlier whose target $y_t$ is already measured. Thus, $\mathbf{e}_t$ reveals which learner currently tracks the regime without future information. The action is the weight vector $\mathbf{a}_t = [w_t^{(1)},\dots,w_t^{(N)}]$.

To guarantee that every action lies on the probability simplex, the actor $\pi_\theta$ outputs the mean $\boldsymbol{\mu}_\theta(\mathbf{s}_t)$ and standard deviation $\boldsymbol{\sigma}_\theta(\mathbf{s}_t)$ of an $(N\!-\!1)$-dimensional Gaussian, and the weights are generated by
\begin{equation}\label{eq:action}
\mathbf{z}_t \sim \mathcal{N}(\boldsymbol{\mu}_\theta,\mathrm{diag}(\boldsymbol{\sigma}_\theta^2)), \quad \mathbf{a}_t = \mathrm{softmax}\big([\kappa \tanh(\mathbf{z}_t),\,0]\big)
\end{equation}
where the zero logit of the reference learner removes the shift-invariance redundancy of softmax, and $\kappa$ bounds the log-ratio between any two weights by $2\kappa$. Two critics $Q_{\psi_1},Q_{\psi_2}$ and their target copies $Q_{\bar\psi_1},Q_{\bar\psi_2}$ evaluate state--weight pairs. Following the maximum-entropy SAC \cite{zhang2023off}, the critics minimize
\begin{equation}\label{eq:critic}
\mathcal{L}_Q = \mathbb{E}\Big[\big(Q_{\psi_i}(\mathbf{s}_t,\mathbf{a}_t) - r_t - \gamma (1-d_t)\, V(\mathbf{s}_{t+1})\big)^2\Big]
\end{equation}
with $V(\mathbf{s}) = \min_{i} Q_{\bar\psi_i}(\mathbf{s},\mathbf{a}') - \alpha \log \pi_\theta(\mathbf{a}'|\mathbf{s})$, $\mathbf{a}' \sim \pi_\theta(\cdot|\mathbf{s})$, and $d_t=1$ when the next sample is not contiguous in time. The actor minimizes $\mathcal{L}_\pi = \mathbb{E}[\alpha \log \pi_\theta(\mathbf{a}|\mathbf{s}_t) - \min_i Q_{\psi_i}(\mathbf{s}_t,\mathbf{a})]$ via the reparameterization trick, and the target critics are updated by $\bar\psi_i \leftarrow (1-\tau)\bar\psi_i + \tau \psi_i$. During deployment, the deterministic action $\mathbf{z}_t=\boldsymbol{\mu}_\theta(\mathbf{s}_t)$ is used.



\subsection{Thermal-Guided Neural Correction}
AIDC power exhibits positive sensitivity to GPU utilization, memory utilization, the number of active jobs, and GPU temperature \cite{saleh2026physics,ji2026sensitivity}: more active streaming multiprocessors draw more dynamic power, and higher junction temperature increases leakage power.

Since the gated recurrent unit (GRU) \cite{cho2014gru} can capture temporal dependencies and extract latent thermal states with few parameters, the thermal sequence $\mathbf{X}_t^{\mathrm{th}} = [u^{\mathrm{gpu}},u^{\mathrm{mem}},T^{\mathrm{gpu}},T^{\mathrm{mem}}]_{t-L+1:t}$ is first fed into a GRU, whose last hidden state is mapped to the residual correction by a fully connected network $g_\phi(\cdot)$:
\begin{equation}\label{eq:corr}
\tilde{y}_{t+h} = \hat{y}_{t+h}^{\mathrm{ens}} + \Delta y_t^{\mathrm{ther}}
\end{equation}
\begin{equation}
\quad \Delta y_t^{\mathrm{ther}} = g_\phi\big(\mathrm{GRU}(\mathbf{X}_t^{\mathrm{th}})\big)
\end{equation}

To preserve the positive sensitivity of power to the dominant drivers, the sensitivities of the correction with respect to the latest GPU utilization and GPU temperature are computed by automatic differentiation:
\begin{equation}\label{eq:grad}
\boldsymbol{\nabla}_t = \left[\frac{\partial \Delta y_t^{\mathrm{ther}}}{\partial u_t^{\mathrm{gpu}}},\; \frac{\partial \Delta y_t^{\mathrm{ther}}}{\partial T_t^{\mathrm{gpu}}}\right]
\end{equation}

The admissible set is the non-negative orthant $\mathcal{C}=\{\boldsymbol{\nabla}\ge \mathbf{0}\}$, whose Euclidean projection is $\Pi_{\mathcal{C}}(\boldsymbol{\nabla}) = \max(\mathbf{0},\boldsymbol{\nabla})$. The thermal-consistency violation is measured by the $\ell_1$ distance between the sensitivity and its projection:
\begin{equation}\label{eq:proj}
\Omega_t = \big\| \boldsymbol{\nabla}_t - \Pi_{\mathcal{C}}(\boldsymbol{\nabla}_t) \big\|_1 = \textstyle\sum_{i\in\{u,T\}} \max(0,-\nabla_{t,i})
\end{equation}
which is zero whenever \eqref{eq:grad} is physically consistent and grows linearly with the violation. Because $\Omega_t$ is itself differentiable with respect to $\phi$ (second-order automatic differentiation), it can be minimized jointly with the forecasting loss.

\subsection{Bayesian Probabilistic Learning and Reward Function}

To quantify forecasting uncertainty under stochastic AIDC loads, a Bayesian output layer is appended to the corrected forecast. Its weights $\mathbf{W}\in\mathbb{R}^{2\times 2}$ are random variables with a mean-field Gaussian posterior $q(\mathbf{W})=\prod_{ij}\mathcal{N}(m_{ij},s_{ij}^2)$ and a standard normal prior $p(\mathbf{W})$ \cite{blundell2015weight}. Given a draw $\mathbf{W}\sim q$, the layer outputs a mean adjustment $\delta_t$ and a scale logit $\rho_t$ through $[\delta_t,\rho_t]=[\tilde y_{t+h},1]\,\mathbf{W}$, and the predictive distribution is:
\begin{equation}\label{eq:bayes}
y_{t+h} \sim \mathcal{N}(\mu_t, \sigma_t^2) \quad \mu_t = \tilde y_{t+h}+\delta_t\;\; 
\end{equation}
\begin{equation}
\sigma_t = \mathrm{softplus}(\rho_t) + \sigma_{\min}
\end{equation}


\begin{algorithm}[!t]
\caption{Offline Training of Proposed TERL}
\label{alg:terl}
\begin{algorithmic}[1]
\STATE Pretrain the LSTM, TCN, and Transformer base predictors by
minimizing mean squared error and retain their best validation checkpoints
\STATE Initialize the actor, twin critics, target critics, thermal
correction module, and variational posterior parameters
\FOR{epoch $=1,\ldots,E$}
    \STATE Construct $\mathbf{s}_t=[\mathbf{X}_t,\mathbf{e}_t]$
    for all training steps
    \STATE Sample $\mathbf{a}_t\sim\pi_\theta(\cdot\mid\mathbf{s}_t)$
    for all training steps 
    \FOR{each shuffled mini-batch $\mathcal{B}$}
        \STATE Compute the ensemble forecast and its thermal correction
        using \eqref{eq:corr}
        \STATE Compute $\Omega_t$ using \eqref{eq:proj} and obtain
        $(\mu_t,\sigma_t)$ for sampled $\mathbf{W}$ using \eqref{eq:bayes}
        \STATE Evaluate $\mathcal{L}_{\mathrm{b},t}$ from the measured
        load $y_{t+h}$ by \eqref{eq:bayesian_loss}
        \STATE Update the base predictors, thermal correction module,
        and variational posterior parameters with Adam by minimizing
        $|\mathcal{B}|^{-1}\sum_{t\in\mathcal{B}}\mathcal{L}_t$ in \eqref{eq:loss}
\STATE Use $r_t$ to update the SAC policy for ensemble forecasting
    \ENDFOR
\STATE Evaluate the model using deterministic
ensemble weights and predictions from $D$ posterior weight draws
\ENDFOR
\end{algorithmic}
\end{algorithm}

For each set of ensemble weights selected by the
SAC policy, the Bayesian output layer maps the thermally corrected ensemble forecast to a predictive distribution. The Gaussian negative log-likelihood term scores the measured load under this distribution, accounting for both the deviation from the predictive
mean and the predictive variance.
This allows SAC to evaluate each weight assignment based on the quality of its probabilistic forecast.
At each
forecast origin $t$, the actor samples an action $\mathbf{a}_t$ from
$\pi_\theta(\cdot\mid\mathbf{s}_t)$ to assign weights to the base predictors. The weighted ensemble forecast is then adjusted by the thermal residual correction in \eqref{eq:corr}. Given a posterior weight sample $\mathbf{W}\sim q(\mathbf{W})$, the Bayesian output layer
maps the corrected forecast $\tilde{y}_{t+h}$ to the conditional Gaussian distribution in \eqref{eq:bayes}, with mean
$\mu_t(\mathbf{W})$ and standard deviation $\sigma_t(\mathbf{W})$.
Using the measured load $y_{t+h}$, the Bayesian forecasting loss is
computed as:
\begin{equation}\label{eq:bayesian_loss}
\begin{aligned}
\mathcal{L}_{\mathrm{b},t}
&=
\mathbb{E}_{q(\mathbf{W})}\!\left[
\log \sigma_t(\mathbf{W})
+
\frac{\big(y_{t+h}-\mu_t(\mathbf{W})\big)^2}
     {2\sigma_t^2(\mathbf{W})}
\right] \\
&\quad+
\beta\,\mathrm{KL}\!\left(
q(\mathbf{W})\,\|\,p(\mathbf{W})
\right)
\end{aligned}
\end{equation}

The expectation gives the Gaussian negative log-likelihood averaged
over the approximate weight posterior, with the additive constant
omitted. It is approximated by Monte Carlo sampling during training.
The coefficient $\beta$ controls the strength of the KL regularization
toward the prior $p(\mathbf{W})$. 

The base predictors, thermal
correction module, and variational posterior parameters are updated by
minimizing the mini-batch average of:
\begin{equation}\label{eq:loss}
\mathcal{L}_t = \mathcal{L}_{\mathrm{b},t}
         + \lambda_{\mathrm{th}}\Omega_t
\end{equation}
where $\Omega_t$ is the thermal sensitivity penalty in \eqref{eq:proj}
and $\lambda_{\mathrm{th}}$ is its weight. 
The SAC reward for the
selected ensemble weights is $r_t = -\mathcal{L}_{t}$.

Consequently, the agent is rewarded for weights that make the \emph{corrected and calibrated} forecast and physically consistent, rather than for weights that only minimize the raw ensemble error.
As summarized in
Algorithm~\ref{alg:terl}, training proceeds in two stages: the LSTM,
TCN, and Transformer predictors are first pretrained by minimizing
the mean squared error and then fine-tuned jointly with the thermal
correction module and Bayesian output layer while the SAC policy
learns to assign ensemble weights. At inference, $D$ Monte Carlo draws $\{\mathbf{W}^{(d)}\}$ give the point forecast $\hat y_{t+h}=\frac{1}{D}\sum_d \mu_t^{(d)}$, and the $(1-\varepsilon)$ prediction interval is formed by the $\varepsilon/2$ and $1-\varepsilon/2$ empirical quantiles of samples $y^{(d)}\sim\mathcal{N}(\mu_t^{(d)},\sigma_t^{(d)2})$ from the resulting Gaussian mixture.

\section{Case Study}
The proposed method is evaluated on the MIT Supercloud dataset \cite{samsi2021supercloud}, which contains large-scale telemetry collected at 1-second resolutions. 
30,000 data instances are split chronologically into 70\% training, 15\% validation, and 15\% testing. The length of the historical time steps is set to 60, and the look-ahead horizon is 1 s. The learning rates for the base predictors and DRL agent are 0.001 and 0.0003. The DRL hyperparameters are set as follows: discount factor $\gamma=0.99$, soft target update coefficient $\tau=0.005$, the replay buffer size $10^5$, the mini-batch size is 256, and the entropy coefficient is automatically tuned. 

Point forecasting accuracy is evaluated by the normalized mean absolute error (nMAE) and normalized root mean squared error (nRMSE), both normalized by the mean training load $\bar y$, e.g., $\mathrm{nMAE}=\frac{1}{|\mathcal{T}|\bar y}\sum_{t\in\mathcal{T}}|y_t-\hat y_t|$. Probabilistic performance is evaluated through the coverage of the 95\% prediction intervals.
The proposed method is compared with: 1) Individual baseline predictors (LSTM, Transformer, and TCN); 2) E-Ensemble: an evenly weighted ensemble of these three models' prediction results.
 All the algorithms are executed on a computer with an Intel Core i9-10900K 3.70 GHz processor and 32 GB of RAM.
 
\subsection{Accuracy Comparison for Various Computing Regimes}
\begin{table*}[!t]
\captionsetup{
font=small,
justification=centering
}
\caption{\textsc{Comparison of Forecasting Error
under Different Computing Scenarios }}
\label{tab:ensemble_necessity}
\centering

\begin{tabular}{|c|cc|cc|cc|cc|}
\hline
\multirow{2}{*}{Method}
& \multicolumn{2}{c|}{Idle}
& \multicolumn{2}{c|}{Busy}
& \multicolumn{2}{c|}{Ramping up}
& \multicolumn{2}{c|}{Ramping down} \\
\cline{2-9}

& nMAE [\%] & nRMSE [\%]
& nMAE [\%] & nRMSE [\%]
& nMAE [\%] & nRMSE [\%]
& nMAE [\%] & nRMSE [\%]\\
\hline

\multicolumn{9}{|c|}{Individual Baseline (* indicates the best performer)}
\\
\hline
LSTM
& 4.28* & 5.21*
& 5.12* & 6.03*
& 8.73 & 10.12
& 8.46 & 9.88 \\
\hline

Transformer
& 4.81 & 5.74
& 5.68 & 6.54
& 7.94* & 9.08*
& 8.11 & 9.37 \\
\hline

TCN
& 6.42 & 7.58
& 5.39 & 6.31
& 10.81 & 12.87
& 7.52*
& 8.74* \\
\hline
\multicolumn{9}{|c|}{Ensemable Learning Model}
\\
\hline
E-Ensemble
& 4.12 & 4.89
& 4.91 & 5.73
& 6.47 & 7.71
& 6.21 & 7.34 \\
\hline

\textbf{Proposed TERL} 
& \textbf{0.82} & \textbf{1.06} 
& \textbf{0.91} & \textbf{1.14} 
& \textbf{1.18} & \textbf{1.39} 
& \textbf{1.05} & \textbf{1.27} \\ 
\hline
\end{tabular}
\end{table*}

\begin{figure}[!t]
\centering
\subfigure[]{
\includegraphics[width=0.92\linewidth]{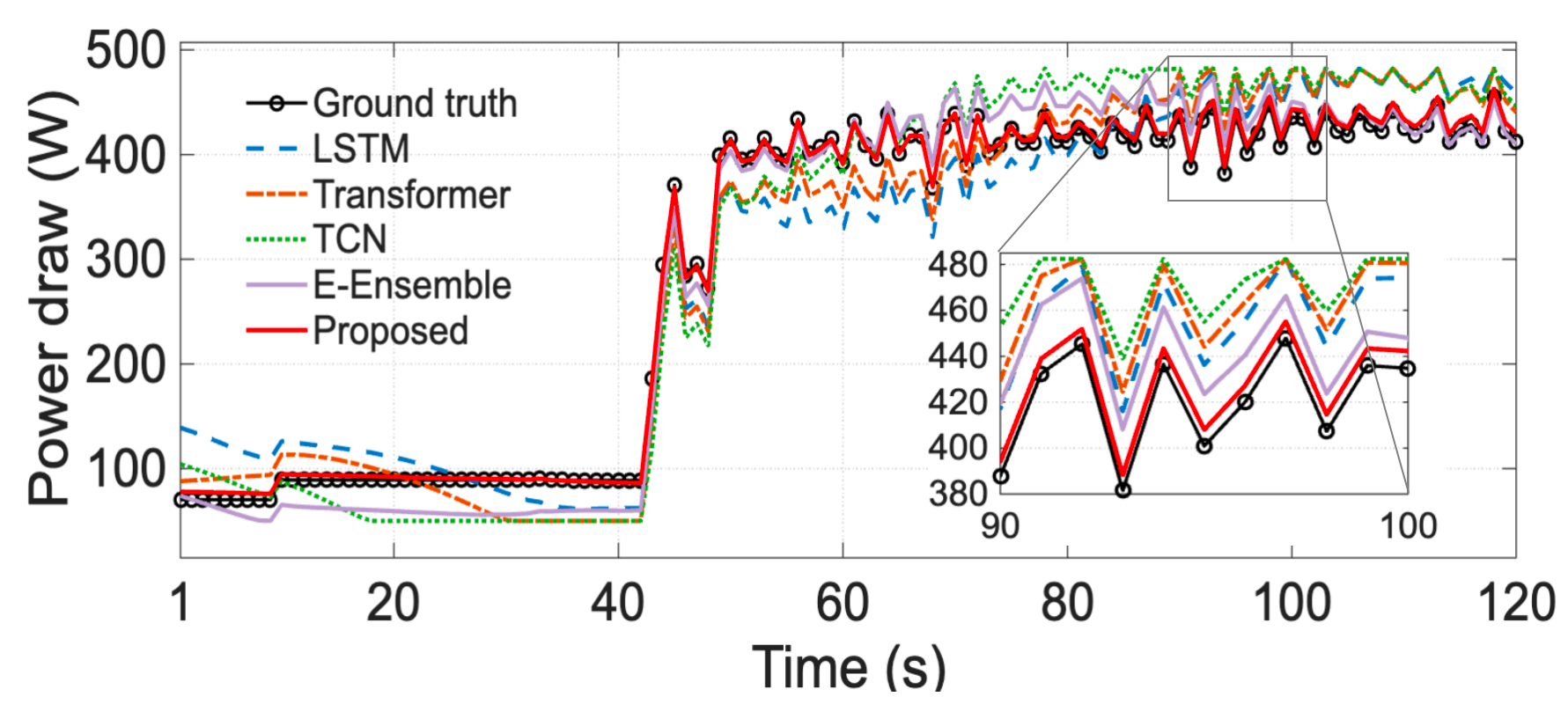}}
\hfill

\subfigure[]{
\includegraphics[width=0.92\linewidth]{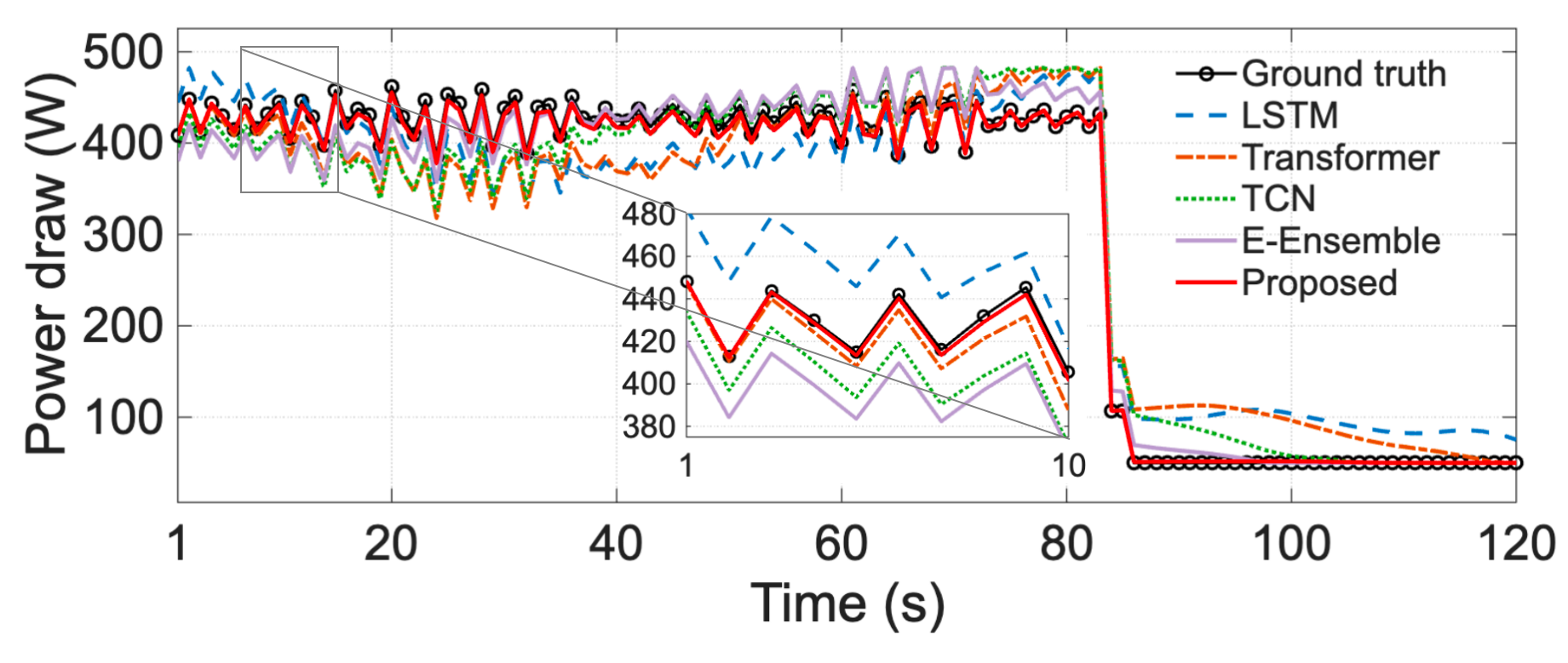}}
\vspace{-9pt}
\caption{Comparison of forecasting results under different computing transitions from (a) idle to ramping up; (b) busy to ramping down.}
\label{fig:9}
\end{figure}

Table~\ref{tab:ensemble_necessity} compares the forecasting performance under computing regimes with recent deterministic deep learning-based methods, where 4500 testing instances at a 1-s sampling interval are grouped into ramping-up, ramping-down, idle, and busy conditions for evaluation. A key observation is that no single base predictor provides the most accurate forecasts across all computing regimes. In contrast to the sensitivity of these methods to rapidly varying regime types, the proposed method consistently achieves the lowest forecasting errors.
Compared with the E-Ensemble method, nMAE are reduced from 6.47\% to 1.18\%, particularly during ramping-up and ramping-down periods with stronger temporal fluctuations. Fig.~\ref{fig:9} further shows that the proposed regime-adaptive physics-guided framework can adaptively capture both stable and rapidly varying workloads.
These results demonstrate that the proposed regime-adaptive physics-guided algorithm can effectively capture both stable and rapidly varying workload dynamics.

In contrast to the sensitivity of these methods to rapidly varying regime types, the proposed method consistently achieves the lowest forecasting errors, with nMAE between 0.82\% and 1.18\% and nRMSE between 1.06\% and 1.39\% in all four regimes. Compared with the E-Ensemble method, nMAE is reduced from 6.47\% to 1.18\% during ramping up and from 6.21\% to 1.05\% during ramping down, i.e., by more than 80\% in the periods with the strongest temporal fluctuations. Moreover, the spread of the proposed nMAE across regimes (0.36 percentage points) is much smaller than that of the individual predictors (3.3-5.4 percentage points), indicating that the adaptive weighting removes the regime dependence of the accuracy. Fig.~\ref{fig:9} further shows two representative transitions. In Fig.~\ref{fig:9}(a), all benchmarks lag behind the step from about 90\,W to 400\,W and then oscillate around the new operating point with biases of up to 50\,W, whereas the proposed forecast follows the step within one sample and tracks the 1-s oscillation pattern. In Fig.~\ref{fig:9}(b), the benchmarks decay slowly to the idle level after the job terminates, while the proposed method immediately settles at the idle power.

\begin{figure}[!t]
   \centering
\includegraphics[width=0.5\textwidth]{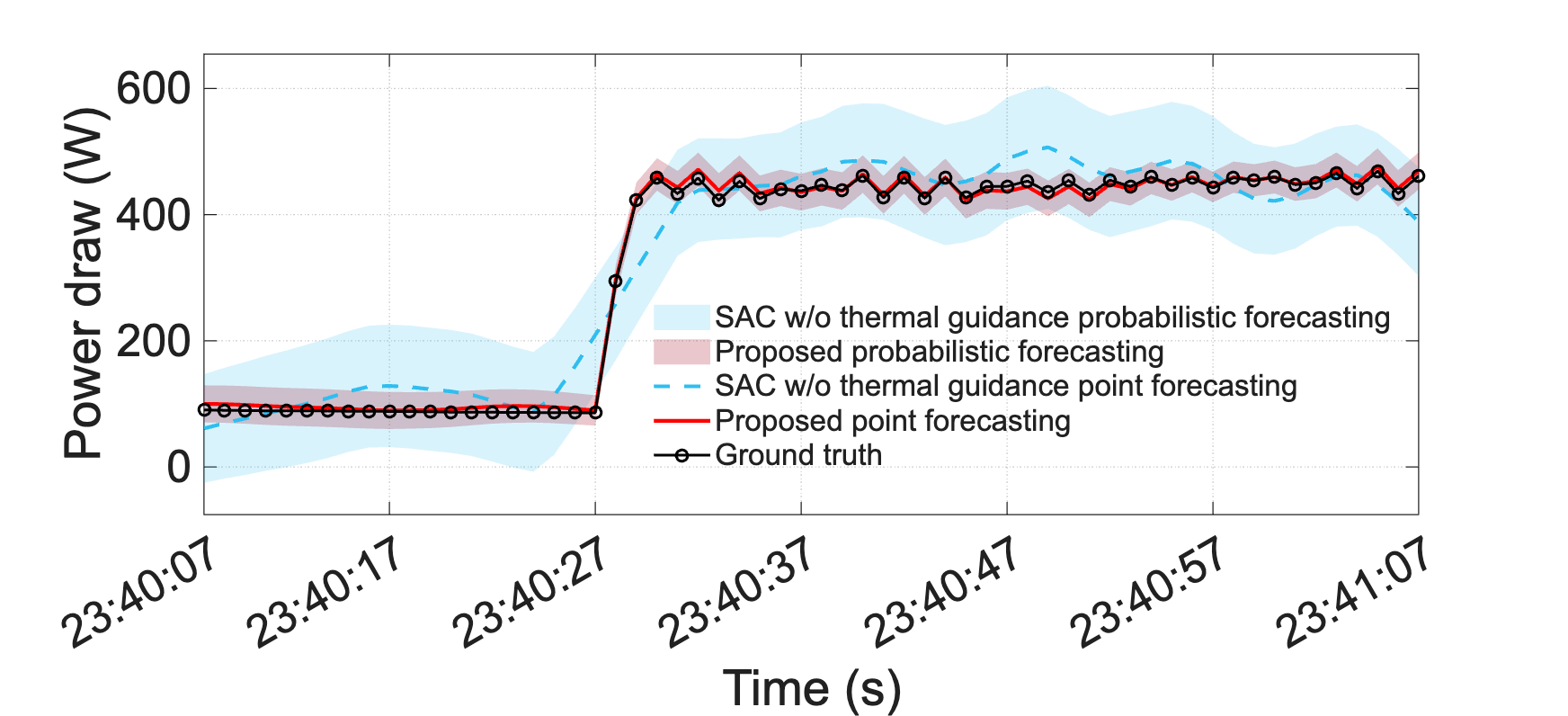}
 \caption{Probabilistic forecasting comparison of the proposed method and SAC-based adaptive weighted ensemble.}
 \label{fig:reward}
 \end{figure}
 \vspace{-5pt}
\subsection{Comparison to DRL Method without Thermal Guidance}
\begin{table}[!t]
\captionsetup{font=small,justification=centering}
\caption{\textsc{Comparison of DRL-Based Adaptive Ensemble Forecasting over the Testing Set}}
\label{tab:drl}
\centering
\begin{tabular}{|c|c|c|c|}
\hline
\makecell{DRL-based\\ensemble learning} & nMAE & nRMSE & \makecell{No. episodes\\to convergence} \\
\hline
SAC \cite{zhang2023off} & 4.35\% & 5.78\% & 730 \\
\hline
\textbf{Proposed} & \textbf{1.32\%} & \textbf{1.37\%} & \textbf{320} \\
\hline
\end{tabular}
\end{table}

To highlight the value of the proposed thermal-guided neural structure and the variational reward, the proposed method is compared with the SAC-based adaptive ensemble, which uses the same base learners and DRL hyperparameters but a vanilla reward without thermal guidance. Table~\ref{tab:drl} shows that the proposed method reduces nMAE from 4.35\% to 1.32\% and nRMSE from 5.78\% to 1.37\%. The draft caption stated a 24-h horizon, which is inconsistent with 4{,}500 testing instances at 1\,s; confirm the evaluation period. Furthermore, the proposed framework converges within approximately 320 training episodes, while the SAC-based weighted ensemble requires around 730 episodes, i.e., a 56\% reduction. This acceleration is attributed to the reward \eqref{eq:loss}: the residual correction absorbs the systematic bias that the ensemble weights cannot remove, and the thermal-consistency term restricts the correction to physically plausible directions, so that the agent receives a smoother and more informative reward signal.

Fig.~\ref{fig:reward} compares the probabilistic forecasts of the two DRL-based methods around a ramping-up event. Without thermal guidance, the SAC-based ensemble produces a point forecast that smears the step over several seconds and a wide interval of about $\pm$100\,W, which even extends below zero power before the step. In contrast, the interval of the proposed method remains narrow in both the idle and busy periods while still enclosing the measured power, because the Bayesian layer learns a small heteroscedastic variance once the corrected ensemble tracks the regime, and widens it only where the measured power oscillates.

\section{Conclusion}
This paper proposes a regime-adaptive TERL algorithm for second-level probabilistic AIDC load forecasting, which couples a SAC ensemble, a sensitivity-projected thermal correction, and a Bayesian output layer through a thermal-consistent variational reward. Case studies on second-level MIT Supercloud telemetry show that, compared with individual deep learning predictors and an evenly weighted ensemble, the proposed method converges faster with narrower prediction intervals than a DRL ensemble without thermal guidance. Future work will extend the thermal constraints to memory utilization and active jobs and use the forecasts for predictive control and job scheduling of grid-tied AIDCs.

\IEEEpubidadjcol

\bibliographystyle{IEEEtran}
\bibliography{IEEEabrv,Citation}

@article{colangelo2025ai,
  title={AI data centres as grid-interactive assets},
  author={Colangelo, Philip and Coskun, Ayse K and Megrue, Jack and Roberts, Ciaran and Sengupta, Shayan and Sivaram, Varun and Tiao, Ethan and Vijaykar, Aroon and Williams, Chris and Wilson, Daniel C and others},
  journal={Nature Energy},
  pages={1--8},
  year={2025},
  publisher={Nature Publishing Group UK London}
}

@article{zhang2023off,
  title={Off-policy deep reinforcement learning with automatic entropy adjustment for adaptive online grid emergency control},
  author={Zhang, Ying and Yue, Meng and Wang, Jianhui},
  journal={Electric Power Systems Research},
  volume={217},
  pages={109136},
  year={2023},
  publisher={Elsevier}
}

@inproceedings{Wang2026RegimeAdaptive,
  author    = {Z. Wang and L. Wang and Y. Zhang and Y. Lin},
  title     = {Regime-Adaptive Weighted Ensemble Learning for
               Computing-Driven Dynamic Load Forecasting in
               AI Data Centers},
  booktitle = {2026 IEEE International Conference on Communications,
               Control, and Computing Technologies for Smart Grids
               (SmartGridComm)},
  year      = {2026},
  address   = {College Station, TX, USA},
  publisher = {IEEE}
}

@INPROCEEDINGS{9916943,
  author={Zhao, Tianqiao and Zhang, Ying and Yue, Meng},
  booktitle={2022 IEEE Power \& Energy Society General Meeting (PESGM)}, 
  title={Scalable Deep Reinforcement Learning-based Volt-VAR Optimization in Distribution Systems: a Mean-field Approach}, 
  year={2022},
  volume={},
  number={},
  pages={1-5},
  doi={10.1109/PESGM48719.2022.9916943}}

@article{abudyak2026mitigating,
  title={Mitigating Data Centers Load Risks and Enabling Grid Support Functions through Grid-Forming Control},
  author={Abudyak, Yousef and Alizadeh, Mohsen and Sun, Wei},
  journal={arXiv preprint arXiv:2602.01013},
  year={2026}
}

@article{choukse2025power,
  title={Power stabilization for AI training datacenters},
  author={Choukse, Esha and Warrier, Brijesh and Heath, Scot and Belmont, Luz and Zhao, April and Khan, Hassan Ali and Harry, Brian and Kappel, Matthew and Hewett, Russell J and Datta, Kushal and others},
  journal={arXiv preprint arXiv:2508.14318},
  year={2025}
}

@article{sheng2026power,
  title={Power for AI Data Centers: Energy Demand, Grid Impacts, Challenges and Perspectives},
  author={Sheng, Yu and Zhang, Chenxuan and Zhu, Zixuan and Xu, Hongyi and Wen, Junqi and Wang, Ruoheng and Yang, Jianjun and Wang, Qin and Bu, Siqi},
  journal={Energies},
  volume={19},
  number={3},
  pages={722},
  year={2026},
  publisher={Multidisciplinary Digital Publishing Institute}
}

@article{kwon2025operational,
  title={Operational Risks in Grid Integration of Large Data Center Loads: Characteristics, Stability Assessments, and Sensitivity Studies},
  author={Kwon, Kyung-Bin and Mukherjee, Sayak and Adetola, Veronica},
  journal={arXiv preprint arXiv:2510.05437},
  year={2025}
}

@article{chen2025electricity,
  title={Electricity demand and grid impacts of AI data centers: Challenges and prospects},
  author={Chen, Xin and Wang, Xiaoyang and Colacelli, Ana and Lee, Matt and Xie, Le},
  journal={arXiv preprint arXiv:2509.07218},
  year={2025}
}

@inproceedings{samsi2021supercloud,
  title={The mit supercloud dataset},
  author={Samsi, Siddharth and Weiss, Matthew L and Bestor, David and Li, Baolin and Jones, Michael and Reuther, Albert and Edelman, Daniel and Arcand, William and Byun, Chansup and Holodnack, John and others},
  booktitle={2021 IEEE High Performance Extreme Computing Conference (HPEC)},
  pages={1--8},
  year={2021},
  organization={IEEE}
}

@article{ji2026sensitivity,
  title={A Sensitivity Analysis of Flexibility from GPU-Heavy Data Centers},
  author={Ji, Yiru and Crozier, Constance and Liska, Matthew},
  journal={arXiv preprint arXiv:2603.27831},
  year={2026}
}

@INPROCEEDINGS{mu2025pesgm,
  author={Mughees, Mariam and Li, Yuzhuo and Chen, Yize and Li, Yunwei Ryan},
  booktitle={2025 IEEE PES General Meeting}, 
  title={Short-Term Load Forecasting for AI-Data Center}, 
  year={2025},
  volume={},
  number={},
  pages={1-5},
  doi={10.1109/PESGM52009.2025.11225317}}

@article{saleh2026physics,
  title={A Physics-Aware Framework for Short-Term GPU Power Forecasting of AI Data Centers},
  author={AlShaikh Saleh, Mohammad and Chawla, Sanjay and Bayhan, Sertac and Abu-Rub, Haitham and Ghrayeb, Ali},
  journal={arXiv preprint arXiv:2605.04074},
  year={2026}
}

@inproceedings{blundell2015weight,
  title={Weight uncertainty in neural networks},
  author={Blundell, Charles and Cornebise, Julien and Kavukcuoglu, Koray and Wierstra, Daan},
  booktitle={Proceedings of the 32nd International Conference on Machine Learning (ICML)},
  pages={1613--1622},
  year={2015}
}

@inproceedings{cho2014gru,
  title={Learning phrase representations using {RNN} encoder-decoder for statistical machine translation},
  author={Cho, Kyunghyun and van Merri{\"e}nboer, Bart and Gulcehre, Caglar and Bahdanau, Dzmitry and Bougares, Fethi and Schwenk, Holger and Bengio, Yoshua},
  booktitle={Proceedings of the 2014 Conference on Empirical Methods in Natural Language Processing (EMNLP)},
  pages={1724--1734},
  year={2014}
}
\let\mybibitem\bibitem

\end{document}